# Multifractal characteristics of optical turbulence measured through a single beam holographic process

**Darío G. Pérez,**[1,*] **Regis Barillé,**[2] **Yohann Morille,**[2] **Sonia Zielińska,**[3] **and Ewelina Ortyl**[3]

[1]*Instituto de Física, Facultad de Ciencias, Pontificia Universidad Católica de Valparaíso (PUCV), Av. Brasil 2950, 23-40025 Valparaíso, Chile*

[2]*LUNAM Université, Université d'Angers/UMR CNRS 6200, MOLTECH-Anjou 2, Bd Lavoisier, 49045 Angers, France*

[3]*Wroclaw University of Technology, Faculty of Chemistry, Department of Polymer Engineering and Technology, 50-370 Wroclaw, Poland*

[*]*dario.perez@ucv.cl*

**Abstract:** We have previously shown that azopolymer thin films exposed to coherent light that has travelled through a turbulent medium produces a surface relief grating containing information about the intensity of the turbulence; for instance, a relation between the refractive index structure constant $C_n^2$ as a function of the surface parameters was obtained. In this work, we show that these films capture much more information about the turbulence dynamics. Multifractal detrended fluctuation and fractal dimension analysis from images of the surface roughness produced by the light on the azopolymer reveals scaling properties related to those of the optical turbulence.

## 1. Introduction

Atmospheric turbulence has a significant impact on the quality of a laser beam propagating through the atmosphere over long distances. Intensity scintillation and beam wandering are the most significant phenomena impacting on the optical quality of such a beam. During half a cen- tury, the boundary layer has received most of the scrutiny of the optical community, but progress in optical communications has drawn attention to study near-to-the-ground propagation. As we get closer to the surface, shear flows are more predominant (rapid changes of tempera- ture and wind direction), and contribute to important deviations from the isotropic Obuhkov- Kolmogorov theory [1–3]. Thus, the measurement of optical quantitates near the ground has become an important issue.

The Obuhkov-Kolmogorov (OK) model is established under the premise that the perturba- tions to the refractive index are scale invariant (self-similar), thus a single exponent, $H = 1/3$, defines all the optical properties; under non-Kolmogorov models this exponent is freed to take other values [4]. This mono-fractal representation is disrupted by the emergence of the inner- and outer-scales in the von Kármán (generalized or not) spectra. Moreover, in turbulence dy- namics the idea of a single exponent controlling the properties of passive scalar, like the refrac- tive index, has been long abandoned; effectively, its real nature is a multi-fractal process [5,6]. Such process is a composition of several sub-structures each with certain local exponent, $h$. That is how von Kármán spectra are just the result of 2-point correlations of this process. Moreover, as the light propagates through the air in turbulent motion, measurements of phase fluctuations, irradiance, or beam wandering inherits some of the multi-fractal properties of the fluctuation of the refractive index [7, 8].

In recent years, a widely adopted method has been used for the determination of the fractal scaling properties, and detection of long-range correlations, in noisy and non-stationary time series: the *multifractal detrended fluctuation analysis* (MF-DFA) [9]. This technique was ex- tended for multi-fractals in higher dimensions by Gu & Zhou [10]: the local exponent is a continuous function of the order $q$ which is associated, roughly speaking, to the size of the fluctuations in the multi-dimensional surface. The value $h$ at $q = 2$ equals the well-known Hurst exponent in the case of homogenous processes—see Sec. 3 for a detailed description of the method. If the local exponent remains constant then we are observing

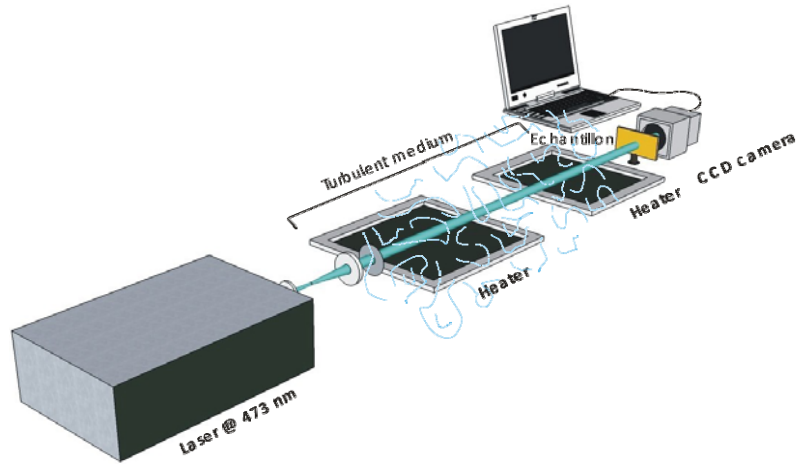

Fig. 1. Experimental set-up to measure the influence of a turbid medium on the laser beam propagation. The heaters are able to modify the refractive index, generating turbulence.

a mono-fractal; that is, all the scales have the same (statistical) self-similarity law.

We have recently reported a new method to characterize the refractive index structure con- stant ($C^2$) [11] by analyzing changes produced by an aberrated wave front incident on the sur- face of an azopolymer thin film. Effectively, the azo molecules on the thin film undergo cyclic *trans-cis* isomerization induced by light absorption; furthermore, a surface pattern develops because of molecular reorientations perpendicular to the polarization direction of the incident light [12]. Moreover, the resulting grating is as regular as stable is the incident light. Because of the turbulence, the laser beam wanders, and then the inscribed grating develops a cluster-like arrangement of patterns. These multi-fractal characteristics should be related, like the structure constant, to the optical properties of the turbulence.

## 2. The experiment

Since convective shear flows present in near-to-the-ground turbulence can be generated in a laboratory by setting up heaters properly, the experimental setup is quite simple. It consists of a collimated laser beam which, after the propagation across the turbulent media, impinges on an azopolymer thin film—Fig. 1.

The polymer films employed in the measurements are made from a highly photoactive azobenzene derivative containing heterocyclic sulfonamide moieties: 3-[{4-[(E)-(4-{[(2,6- dimethylpyrimidin-4-yl) amino] sulfonyl}phenyl) diazenyl]phenyl}-(methyl)amino]propyl 2- methylacrylate. These were prepared by spin coating azobenzene (diluted in THF with a con- centration of 50mg/ml) on a glass substrate. The average thickness is around 1 m, and its absorbance has a maximum value of 1.9 at 438 nm—for more information see [13]. A coherent and linearly polarized light induces the formation of surface relief gratings on the azopolymer films; this phenomenon results from isomerization-induced translation in which the molecules migrate almost parallel to the polarization direction. Additionally, the surface gratings can be easily removed by using a circularly polarized laser beam or by heat treatment, thus making these films highly appropriate for repetitive measurements.

Therefore, for the purpose of registering the beam statistics into the film, a 10 mW diode- pumped solid-state (DPSS) laser emitting at 473 nm is used to maximize the absorption of the azopolymer (the absorbance is 1.6). It is collimated, by an afocal Keplerian telescope, into a beam of 5.1 mm in diameter by interposing a pupil in the central part of it, a quasi-plane wave with a diameter of 3.1 mm is prepared to travel through the turbulence. The convective turbulence, extended over a one meter region, is produced by two electric hot plates (at 98°C) and a hot stream of air (42°C, with mean flow speed between 5.12 and 7.6 m/s controlled by a hose of varying diameter). It was characterized, before the actual experiment, by measuring the beam centroid wandering with a CCD camera [14]—the sensor size is 8.4 ×6.2 mm (JAI-Pulnix monochrome). The camera was located sufficiently away from the end of the turbulent region to improve the resolution of lateral displacements. Four different experimental conditions were produced: the inner-scale has been found ranging from 1.7 mm to 2.6 mm, while the structure constant, $C_n^2$ went from $9.4 \times 10^{-13}$ to $4.1 \times 10^{-12}$ m$^{-2/3}$. These values correspond to weak fluctuation conditions, since the estimated Rytov variance results $\sigma_R^2 \ll 1$ for the propagation parameters given.

Once each experiment starts, the laser beam statistics is registered on the azopolymer thin film over a period of about 9 minutes. The patterning process builds up by an average irradiance between 100 and 300 mW/cm$^2$, but most of radiation is not absorbed. Behind the film a diffraction image is produced because of the polymer surface rearrangements, and then the temporal evolution of the first-order diffraction is observed in the polymer surface rearrangements, and then the temporal evolution of the first-order diffraction is observed in a CCD camera, when the diffraction intensity saturates the laser beam is switched off, and then the thin film is removed.

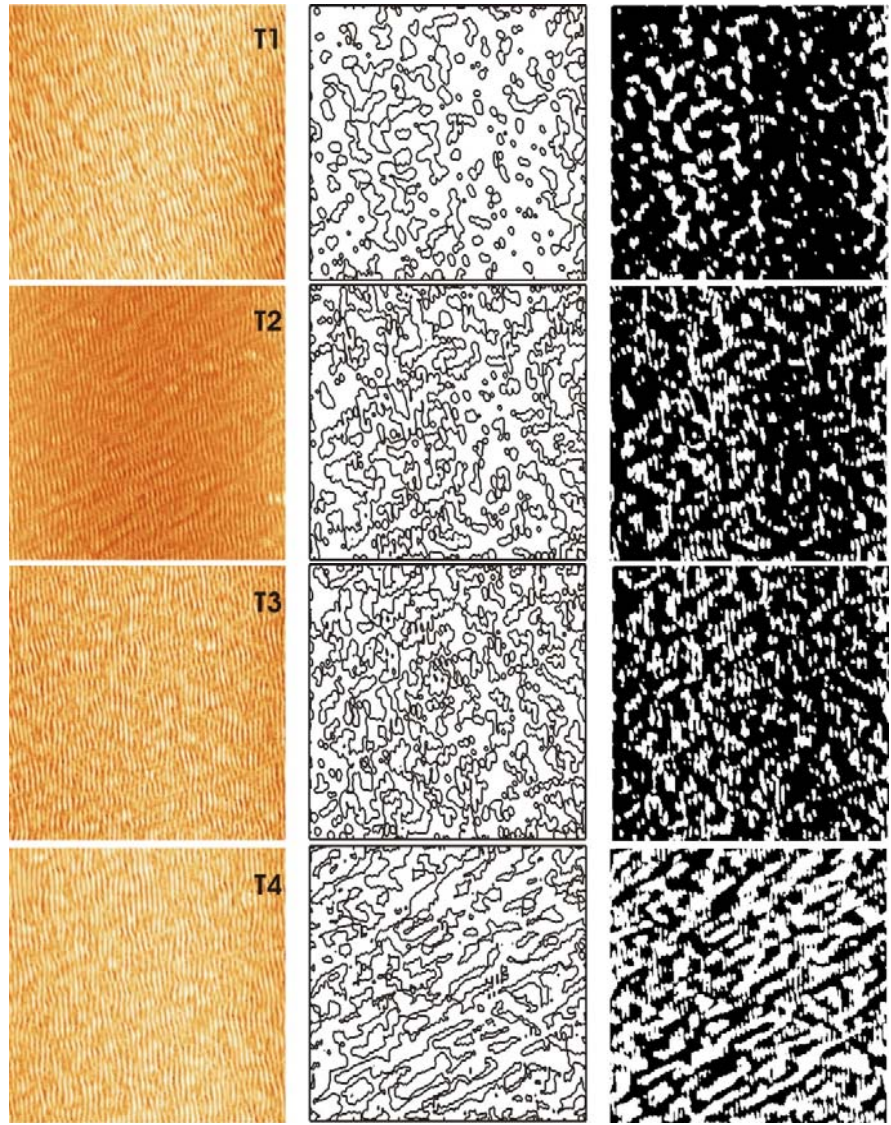

Fig. 2. Surface patterns on the materials obtained for four different turbulent conditions (left). The central patterns correspond to the contours of the Gaussian surfaces by thresholding the height of the topography with a fixed spacing $\Delta$ between the height of the successive contours (center). Filling of the central patterns corresponding to the contours of the gaussian surfaces (right). The four patterns corresponds to $C_n^2 = 9.4 \times 10^{-13}$, $1.4 \times 10^{-12}$, $3.8 \times 10^{-12}$, and $4.1 \times 10^{-12}$ m$^{-2/3}$.

As has been reported before, fractal characteristics of the wave front should appear from inspecting these clustering patterns. We implemented two approaches to scrutinize this property.

### 3. The analysis

The surface patterns inscribed on the thin films were imaged with an atomic force microscope. Figure 2 (left) shows the topography of the thin-film surface after the laser inscription for four different turbulent conditions.

For these turbulences, the self-organized patterns have an average grating pitch of 800 ± 30 nm, yet separated domains are observed as a consequence of the random angle-of-arrival fluctuations of the beam. These domains on the surface are separated by boundaries where the periodic structures almost vanish. The occurrence of these domains varies depending on the turbulence strength, while the high-frequency periodic patterns persist for any intensity level.

We first estimated the (box-counting) fractal dimension for the sets obtained at each turbu- lence intensity. To reduce the influence of the high-frequency periodic pattern, we applied a Gaussian low-pass filter Afterwards, for each turbulence intensity, we prepare sets of contour domains by thresholding at an intensity level high enough to eliminate any residual periodicity. This threshold is determined by the empirical value $\Delta = I_{avg}/\sqrt{6}$, where $I_{avg}$ is the mean value of the surface roughness.

This results in outline drawings consisting of non-intersecting closed level curves, Fig. 2 (center), and then each one of these contours is filled to create black and white image masks, Fig. 2 (right). The fractal dimension is a measure of the space-filling properties of a given characteristics set. The box-counting, or capacity, dimension $D_c$ is one of several definitions of fractal dimension—see [15] for a detailed discussion on alternate def- initions. It is calculated by counting the number of boxes of side length $\delta$ covering the level contour set, $N(\delta)$; thus, the quotient $-\log N(\delta)/\log \delta$ will approach a constant scaling value $D_c$ as the refinement length $\delta$ becomes smaller [15]. Figure 3 shows the relation between the capacity dimension of the azopolymer thin film and the turbulence strength. Two different regimes can be observed depending on the structure constant value, and a linear fit is easily obtained in each case $D_c = a.(C_n^2) + b$; $a = 4.78 \times 10^{11}$ m$^{-2/3}$ and $b = 1.65$ for $< 1.5 \times 10^{-12}$ m$^{-2/3}$, and $a = 1.81 \times 10^{11}$ m$^{-2/3}$ and $b = 1.92$ for $\geq 1.5 \times 10^{-12}$ m$^{-2/3}$.

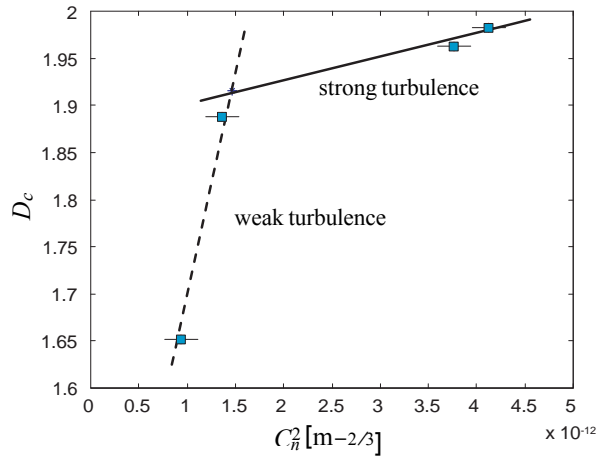

Fig. 3. Box-counting dimension of the patterns as a function of the refractive index structure constant $C_n^2$.

The box-counting dimension is a property due to second order statistical correlations in the original random field (the random wavefront phase); therefore, higher- or lower-order (fractional) moments should give us a more complete understanding of the correlations present in the propagated light, and registered by the thin film. Without turbulence, the patterns are periodic and thus fractal behaviour is absent on the film; however, the presence of turbulence produces spatial variations on the pattern, and diverse scale invariant structures develop.

A *two-dimensional* multifractal detrended fluctuation analysis on a thin-film pattern struc- ture, Fig. 2(left), should reveal these fractal, and long-term, correlations [9, 10, 16–19]. This technique requires of three steps briefly sketched here (following the procedure given in [10]): an image array $X$, with $N \times M$ elements, is partitioned into $M_S \times N_S$ disjoint sub-images (square regions) of size $s \times s$, where $N_S = [N/s]$ and $M_S = [M/s]$. For each sub-image $X_{v,w}$, the cumulative sum is:

$$u_{v,w}(i,j) = \sum_{m=1}^{i}\sum_{n=1}^{j} X_{v,w}(n,m) - <X_{v,w}>$$

—$(v,w) \in M_S \times N_S$ and $(i,j) \in s \times s$. Each cumulative sum, $u_{v,w}$, has a local trend that can be estimated by fitting with a bivariate polynomial $\tilde{u}_{v,w}$—estimated through the least square method. Thus, a new array image is obtained by removing this trend from the original image: $\varepsilon_{v,w} = u_{v,w} - \tilde{u}_{v,w}$. The variance for each sub-image $(v,w)$, for a partition of size $s$, is:

$$F^2(v,\omega,s) = \frac{1}{s^2}\sum_{i,j=1}^{s,s} \varepsilon_{v,\omega}(i,j)^2$$

finally, the $q^{th}$-order fluctuation function is calculated as:

$$F_q(s) = \left\{\frac{1}{M_S N_S}\sum_{v,\omega=1}^{M_S N_S}[F^2(v,\omega,s)]^{q/2}\right\}^{1/q}$$

This fluctuation function reveals the power-law scaling for large values of $s$ on these irregular sets; that is, $F_q(s) \sim s^{h(q)}$ [10]. Therefore, the *generalized Hurst exponent* $h(q)$ can be estimated from linear fitting log-log plots for large enough $s$.

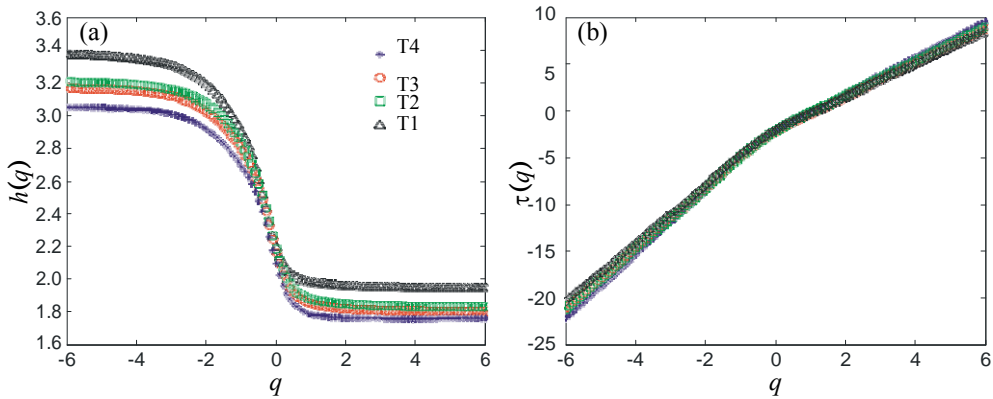

Fig. 4. The Hurst (a) and multifractal scaling (b) exponents as functions of the $q^{th}$ moment for different turbulence intensities.

The generalized Hurst exponent shown in Fig. 4(a) was obtained from the turbulent conditions producing the patterns observed in Fig. 2 after the procedure detailed above was applied. In all these examples, $h(q)$ strongly depends on $q$; therefore, for all the intensities produced, the turbulences give multifractal patterns on the thin films. Yet, the computation of the Rényi exponent $\tau(q)$, also known as the multifractal scaling exponent, reveals more about the multifractal characteristics of these patterns—see Fig. 4(b). This exponent is related to the generalized Hurst exponent through $\tau(q) = qh(q) - 2$, and it represents the spatial structure of the image as a function of the moment $q$. In fact, as discussed in [20], the lowest moments are related to: the capacity of the support of the analyzed function ($-\tau(0)$, assuming is nowhere differentiable), the capacity of the function graph (max $\{2, 2 - \tau(1)\}$, for a two-dimensional function), or the scaling exponent of the spectral density ($\beta = \tau(2) + 3$). The box-counting dimension of the complete image (graph) results an increasing function of the structure function as with the capacity of the image level sets, giving: 2.02 (T1), 2.12 (T2), 2.15 (T3), and 2.22 (T4).

Roughly speaking, the Hölder exponent $\alpha$, or singularity strength, measures the local scaling of a continuous function. The singularity spectrum $f(\alpha)$ gauges the fractal dimension of the set of points having a given $\alpha$ singularity strength—see [15] for a precise definitions. We can relate $\alpha$ and $f(\alpha)$ to $h(q)$ with the Legendre transform:

$$\alpha = h(q) + q\frac{dh(q)}{dq} \text{ and } f(\alpha) = q[\alpha - h(q)] + 1 = \alpha.q(\alpha) - \tau(q(\alpha))$$

The width of the spectra $f(\alpha)$ is a measure of how multifractal is the signal; usually, the mul- tifractality is characterized by the difference between the maximum and minimum values of the singularity strength, $\alpha_{max} - \alpha_{min}$. The singularity spectrum for a monofractal image corresponds to a Dirac delta function, there is one value of $\alpha$; otherwise, it is a distribution of Hölder exponents. The wider the spectrum is around its maximum, the higher is the heterogeneity of analyzed structures. Figure 5 displays the singularity spectrum and maxima for the four rep- resentative temperatures discussed before (T1, T2, T3, and T4). The maxima show the same trend observed for the capacity dimension estimated by box-counting, $D_c$, and the one evaluated through $\tau(1)$. As the intensity of the turbulence grows the singularity spectrum moves forwards towards higher exponents, although the multifractality richness remains constant because the width is unchanged by the state of the turbulent flux.

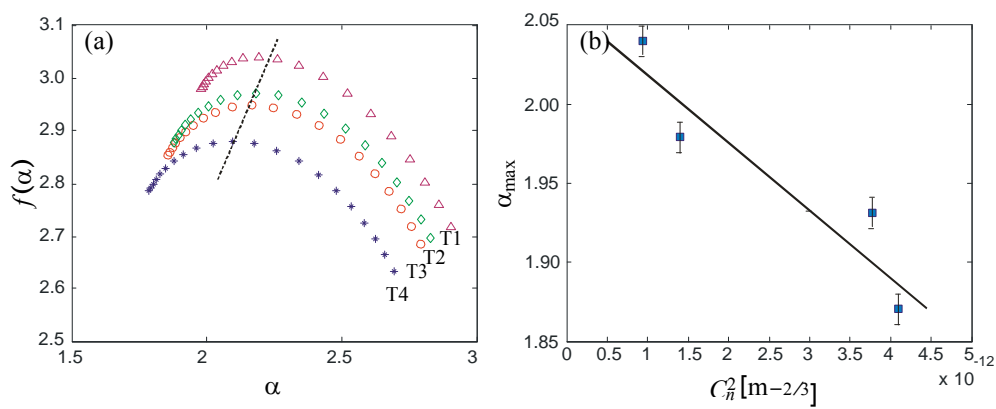

Fig. 5. (a) Singlularity spectrum $f(\alpha)$ as a function of the Hölder exponent for different turbulences estimated from the thin-film patterns. (b) Maximum of the singularity spectrum as a function of the structure constant of the refractive index $C_n^2$

## 4. Conclusions

In this work, we have shown that the inherent multifractal nature of the turbulence is engraved on the azompolymer thin film by the propagated laser beam. More precisely, this new method measures changes in the fractality of the wavefront as the convective turbulence develops— measured by its strength. mension of a particular iso-level curve at $I_{avg}/\sqrt{6}$. Their fractal dimension increases with the intensity of refractive index structure constant. This phenomenon have already been reported for numerical simulation of irradiance distribution [7]. To our knowledge, this work is the first experimental confirmation of this behaviour.

Yet, it is well-known that the usual box-counting technique is prone to errors in finding the correct values to the dimension; therefore, we introduce a bidimensional multifractal de- trended fluctuation analysis to understand further the fractal characteristics recorded by the erratic movement of the laser beam over the thin film. The generalized Hurst exponent shows a strong multifractal behaviour, independent of the turbulence intensity: the convexity and inflec- tion of the curve with respect to the $q^{th}$ moment remains unchanged. Although, the whole curve moves down with the increasing strength of the turbulence (measured by $C_n^2$ ). This changes the values of the Hurst and the corresponding multifractal scaling exponents, Fig. 4. Furthermore, decreasing values of $h(2)$ with increasing strength have been reported recently for convective turbulence—with values above the classical 5⁄6 Hurst exponent [8].

In particular, the capacity of the image can be estimated from $\tau(1)$. As discussed earlier, it grows with the structure constant. Notably, it approaches faster than $h(2)$ to the value expected from the Kolmogorov model applied to the wave-front phase [4]: $D_C = 3 - 5/6 \cong 2.17$. This result seems to indicate that turbulent active regions tend to organize into the fractal structure of classical isotropic turbulence before the second order correlations measured by $h(2)$ reach that limit.

On the other hand, the singularity spectra attain their maxima at higher Hölder exponents than the generalized Hurst values, but the shape of the spectra is mostly preserved; particu- larly, the width of these spectra remains the same. Also, we observe that most Hölder expo- nents are located to the right of the maxima. This may be the reason behind the box-counting dimension technique having quite different values compared to, the more precise, capacity di- mension through MF-DFA [21]. Undoubtedly, because of the wide singularity spectra usual box-counting methods of applied to iso-level curves of these gratings should be avoided.

Finally, this work extents our previous work [11]; not only the turbulence strength is registered in the azopolymer thin films by the laser beam. Indeed, long-term multifractal characteristics of the wavefront are discovered in these gratings.

**Acknowledgments**

RB, YM, SZ and EO have been supported by CNRS through a collaborative PICS project with the Wroclaw University of Technology. DGP has been supported by Comisión Nacional de Investigación Científica y Tecnológica (CONICYT, FONDECYT Grant No. 1140917, Chile), partially by Pontificia Universidad Católica de Valparaíso (PUCV, Grant No. 123.731/2014, Chile).